\documentclass[]{spie}  

\usepackage{amsmath,amsfonts,amssymb}
\usepackage{graphicx}
\usepackage[colorlinks=true, allcolors=blue]{hyperref}
\usepackage{float}

\newcommand{\scoob}{SCoOB }

\title{The space coronagraph optical bench (SCoOB): 3. Mueller matrix polarimetry of a coronagraphic exit pupil}

\author[a,b]{Jaren N. Ashcraft}
\author[b]{Ewan S. Douglas}
\author[b]{Ramya M. Anche}
\author[b]{Kyle Van Gorkom}
\author[a]{Emory Jenkins}
\author[c]{William Melby}
\author[c]{Maxwell A. Millar-Blanchaer}

\affil[a]{James C. Wyant College of Optical Sciences, University of Arizona, 933N Cherry Avenue, Tucson, Arizona, 85721, USA}
\affil[b]{Steward Observatory, University of Arizona, 933N Cherry Avenue, Tucson, Arizona, 85721, USA}
\affil[c]{Department of Physics, University of California, Santa Barbara, CA, 93106, USA}

\authorinfo{Further author information: (Send correspondence to J.N.A.)\\J.N.A.: E-mail: jashcraft@arizona.edu}

\begin{document} 
\maketitle

\begin{abstract}
High-contrast imaging in the next decade aims to image exoplanets at smaller angular separations and deeper contrasts than ever before. A problem that has recently garnered attention for telescopes equipped with high-contrast coronagraphs is polarization aberration arising from the optics. These aberrations manifest as low-order aberrations of different magnitudes for orthogonal polarization states and spread light into the dark hole of the coronagraph that cannot be fully corrected. The origin of polarization aberrations has been modeled at the telescope level. However, we don't fully understand how polarization aberrations arise at the instrument level. To directly measure this effect, we construct a dual-rotating-retarder polarimeter around the \scoob high-contrast imaging testbed to measure its Mueller matrix. With this matrix, we directly characterize the diattenuation, retardance, and depolarization of the instrument as a function of position in the exit pupil. We measure the polarization aberrations in the Lyot plane to understand how polarization couples into high-contrast imaging residuals.  
\end{abstract}


\section{INTRODUCTION}
Polarization aberrations are a known limiter to high-contrast imaging in the next decade. In advance of the Astro 2020 decadal survey, the LUVOIR and HabEx studies imposed constraints on the design of their OTA in order to minimize the presence of polarization aberrations that ultimately reach their coronagraphs\cite{gaudi2020habitable, Will_polarization_luvoir}. The next-generation giant segmented mirror telescopes (GSMTs) experience considerable polarization aberration due to the large change in the angle of incidence in the telescope to fit within a reasonably-sized dome \cite{anche2023polarization}. Much work has been done to model the polarization aberrations in astronomical telescopes; however, little is understood about the polarization aberrations at the instrument level.

To achieve $10^{-10}$ contrast on a space-borne coronagraph, similar to the one on the future Habitable Worlds Observatory (HWO), the telescope will require unprecedented stability and performance. Quantifying all possible sources of error is necessary in order to achieve this target contrast. Inspired by the CDEEP mission concept\cite{Maier2020}, the Space Coronagraph Optical Bench (\scoob)\cite{Ashcraft2022,VanGorkom2022,VanGorkom2024, Anche2024} is a vacuum-compatible coronagraph testbed at the University of Arizona's Space Astrophysics Laboratory aimed at prototyping high-contrast imaging technology for space-based telescopes. \scoob was designed to achieve high static contrast with relatively low polarization aberration\cite{Maier2020}, but the polarization aberrations of the coronagraph have never been evaluated. In this study, we construct a polarimeter around \scoob to understand the presence of polarization aberrations at the coronagraph level.

Measurement of the polarization aberrations of a system requires an understanding of how it transforms the polarization state across a beam. Polarization aberrations are typically computed from the Jones pupil, which is a spatially varying Jones matrix at the exit pupil of an optical system. However, Jones matrices cannot be measured directly. In contrast, the Mueller pupil is a quantity that can be evaluated with a series of power measurements. This loses track of the global phase of the polarization aberration but enables an evaluation of how the polarization state is aberrated with respect to a reference phase. In this study, we seek to measure the Mueller pupil of \scoob to gain an understanding of the degree of polarization aberration present in coronagraphic instruments. In Section \ref{sec:theory}, we outline the principle of a dual-rotating-retarder Mueller polarimeter (DRRP) that we use to measure the \scoob Mueller matrix. In Section \ref{sec:assembly}, we outline the assembly, calibration, and validation of the DRRP in the UA Space Astrophysics Laboratory. In Section \ref{sec:results}, we show the results of placing the polarimeter around \scoob to measure the Mueller pupil of the coronagraph. In Section \ref{sec:discussion} we evaluate what the results mean for calibrating \scoob's static contrast floor, and outline a path for future investigation.

\section{Dual-rotating Retarder Mueller Matrix Polarimetry}
\label{sec:theory}
The DRRP is a commonly used device for measuring the Mueller matrix of an instrument. The DRRP was first described in Azzam 1978\cite{Azzam:78}, where they describe the determination of a Mueller matrix by Fourier analysis of the detected signal from a device that modulates the polarization state sent to and observed from a unit under test. Smith 2002\cite{Smith:02} describes the optimization of the free parameters of the DRRP for fast and precise measurement of the Mueller matrix. Chapter 7 of Chipman, Lam, and Young\cite{CLY} describes a variation on the data reduction for a DRRP that is computationally simple to implement, which we reproduce in this section.

The Mueller matrices for a linear polarizer ($\mathbf{LP}$) and linear retarder ($\mathbf{LR}$) are given by Equations \ref{eq:linear_polarizer} and \ref{eq:linear_retarder}. 
\begin{equation}
    \mathbf{LP(\theta)} = 
    \begin{pmatrix}
        1 & cos(2\theta) & sin(2\theta) & 0 \\
        cos(2\theta) & cos^{2}(2\theta) & cos(2\theta)sin(2\theta) & 0 \\
        sin(2\theta) & cos(2\theta) & sin^{2}(2\theta) & 0 \\
        0 & 0 & 0 & 0 \\
    \end{pmatrix}
    \label{eq:linear_polarizer}
\end{equation}
\begin{equation}
    \mathbf{LR(\theta,\delta)} =
    \begin{pmatrix}
        1 & 0 & 0 & 0 \\
        0 & cos^{2}(2\theta) + cos(\delta)sin^{2}(2\theta) & (1-cos(\delta))cos(2\theta)sin(2\theta) & -sin(\delta)sin(2\theta) \\
        0 & (1-cos(\delta))cos(2\theta)sin(2\theta) & cos(\delta)cos^{2}(2\theta) + sin^{2}(2\theta) & cos(2\theta)sin(\delta) \\
        0 & sin(\delta)sin(2\theta) & -sin(\delta)cos(2\theta) & cos(\delta )\\
    \end{pmatrix}
    \label{eq:linear_retarder}
\end{equation}

\begin{figure}
    \centering
    \includegraphics[width=\textwidth]{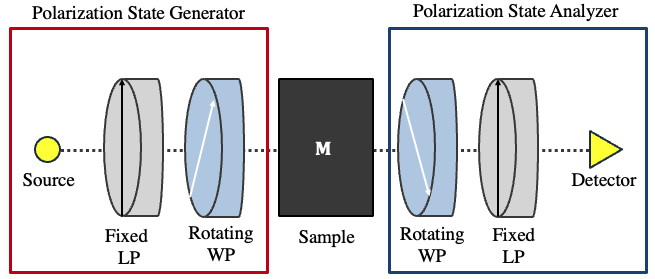}
    \caption{Diagram illustrating the principle of a DRRP. Starting from some source, light propagates through a fixed linear polarizer (LP) and then a rotating waveplate (WP) that forms the polarization state generator, which modulates the light sent to the sample ($\mathbf{M}$). Light from the sample under test then passes through another rotating WP and fixed LP, which modulates the polarization state that the final analyzer sees. This encodes analysis of the polarization state into a series of intensity measurements observed by a detector, which are of a fixed polarization state.}
    \label{fig:drrp_diagram}
\end{figure}

A DRRP can be characterized by a sequence of a source, polarization state generator ($\mathbf{PSG}(\theta_{p,1},\theta_{r,1},\delta_{r,1})$, given by Equation \ref{eq:psg}), a sample to measure, and a polarization state analyzer ($\mathbf{PSA}(\theta_{p,2},\theta_{r,2},\delta_{r,2})$, given by Equation \ref{eq:psa}). A diagram of the DRRP is shown in Figure \ref{fig:drrp_diagram}.

\begin{equation}
    \mathbf{PSG} = \mathbf{LR}(\theta_{r,1},\delta_{1}) \mathbf{LP}(\theta_{p,1})
    \label{eq:psg}
\end{equation}
\begin{equation}
    \mathbf{PSA} = \mathbf{LP}(\theta_{p,2}) \mathbf{LR}(\theta_{r,2},\delta_{2})
    \label{eq:psa}
\end{equation}

The DRRP measures the Mueller matrix of a sample $\mathbf{M}$ through a series of measurements where the retarders are progressively rotated. In this study, for every $n^{\circ}$ that the polarization state generator waveplate is stepped, the polarization state analyzer is stepped by $5n^{\circ}$. However, Smith\cite{Smith:02} notes that other angular velocity ratios may be more optimal for rapid polarimetry.

Using a Mueller matrix model of the polarimeter, we construct a polarimetric data reduction matrix $\mathbf{W}$ that is given by Equation \ref{eq:kron},

\begin{equation}
    \mathbf{W} = \mathbf{psa} \otimes \mathbf{psg},
    \label{eq:kron}
\end{equation}

where $\otimes$ is the Kronecker product, and $\mathbf{psg},\mathbf{psa}$ are the first row and column vectors of the $\mathbf{PSG}$ and $\mathbf{PSA}$, respectively. The Mueller matrix is computed using the pseudo-inverse of $\mathbf{W}$, $\mathbf{W}^{+}$, multiplied by the vector of observed power $\mathbf{p}$, as shown in Equation \ref{eq:mueller_data_reduce},

\begin{equation}
    \mathbf{m}_{meas} =  \mathbf{W}^{+} \mathbf{p}
    \label{eq:mueller_data_reduce}
\end{equation}

where the result $\mathbf{m}_{meas}$ is a $16 \times 1$ vector that contains the Mueller matrix.

\section{Assembly and Operation of a Dual Rotating-Retarder Mueller Polarimeter}
\label{sec:assembly}
We construct the DRRP facility in the UA Space Astrophysics Laboratory (UASAL) using entirely commercial off-the-shelf components, a list of which is included in the appendix of this manuscript. Figure \ref{fig:lab_setup} shows the DRRP assembled in the clean tent in UASAL. We use the Newport Agilis Piezo Motor Driven Rotation Stages with the AG-UC8 controller to rotate the quarter-wave retarders used in the DRRP. These rotation stages are driven via serial communication and lack encoders that indicate their current angular position, so we must calibrate the step sizes of the stages. To provide an open-source platform for serial communication of these rotation stages, we have created the \verb|katsu| Python package\cite{Ashcraft_Katsu}. \verb|katsu| utilizes the \verb|Pyserial| Python package for serial communication with a more user-friendly API than the standard command string interface supplied by Newport\footnote{\url{https://www.newport.com/mam/celum/celum_assets/np/resources/Agilis_Piezo_Motor_Driven_Components_User_Manual.pdf?1}}, and also contains routines for Mueller matrix modeling, polarimetric data reduction, and Mueller matrix analysis. We use \verb|katsu| for the entirety of the investigations contained in this proceedings.

\begin{figure}
    \centering
    \includegraphics[width=0.6\textwidth]{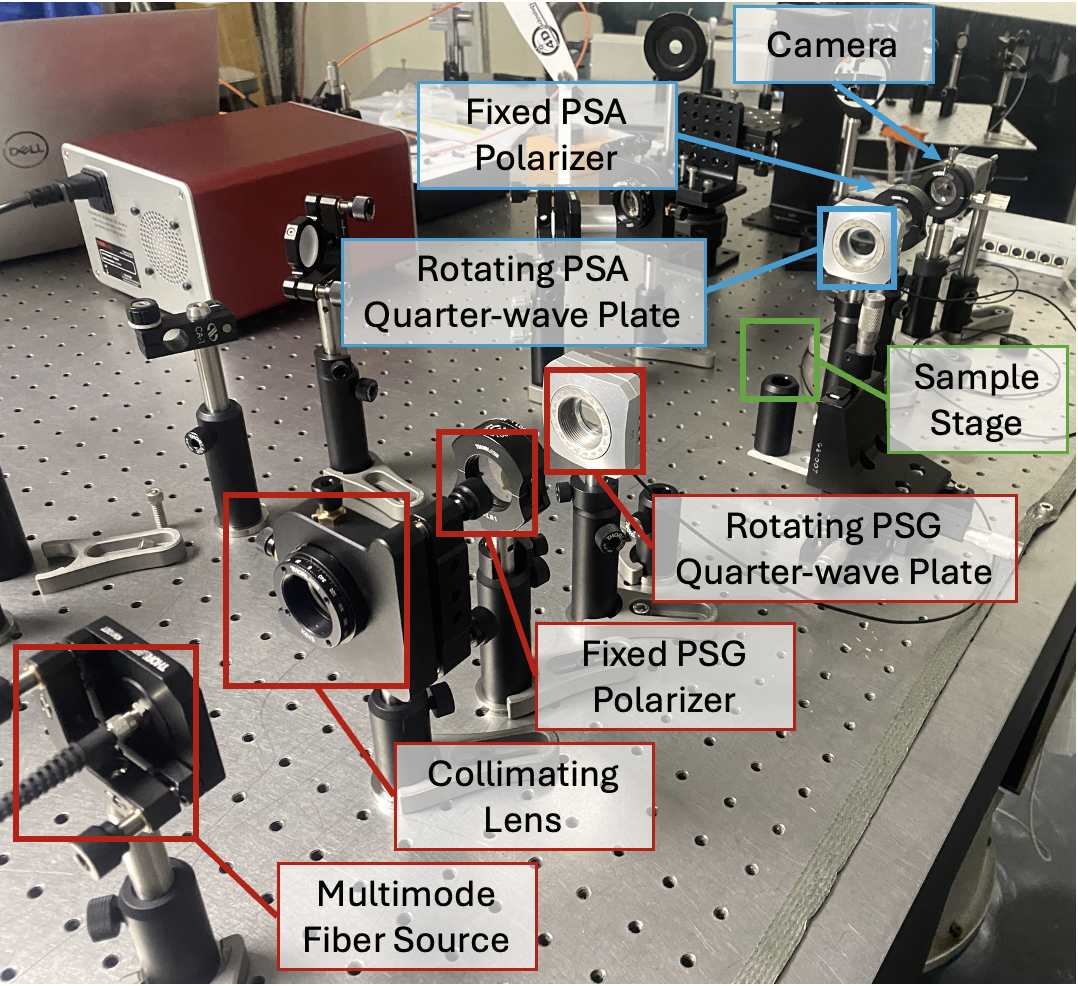}
    \caption{The DRRP facility in the UA Space Astrophysics Laboratory. The polarization state generator optics are highlighted in red. The polarization state analyzer optics are highlighted in blue. Shown in green is a translation stage that an optic under test can be mounted in.}
    \label{fig:lab_setup}
\end{figure}



\subsection{Polarimeter Calibration}
The goal of Mueller polarimeter calibration is to determine the most accurate polarimetric data reduction matrix $\mathbf{W}$ as possible. To begin the calibration, we first perform an air measurement. We know the Mueller matrix of air to be nearly an identity matrix, which constrains the state of our system. The Mueller matrix of the polarimeter for a given state is then given by Equation \ref{eq:nom_state},

\begin{equation}
    \mathbf{M}_{cal} = \mathbf{PSA}(\theta_{p,2},\theta_{r,2},\delta_{2})\mathbf{PSG}(\theta_{p,1},\theta_{r,1},\delta_{1}).
    \label{eq:nom_state}
\end{equation}

\begin{figure}
    \centering
    \includegraphics[width=\textwidth]{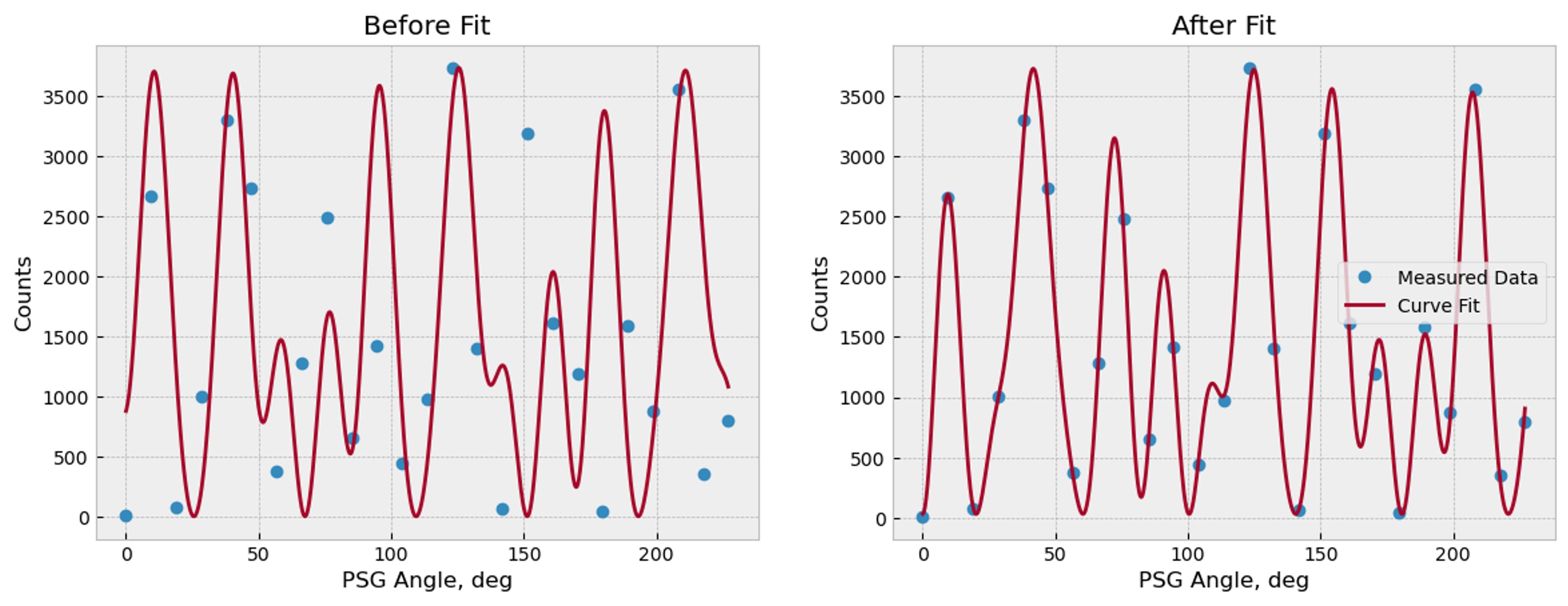}
    \caption{(Left) The measured mean power at the detector plane (shown as blue dots) and the expected model of the power before fitting (shown as a red line). (Right) The measured mean power and expected model after fitting to the 8 free parameters of the polarimeter. This calibration step constrains our Mueller matrix model of the polarimeter.}
    \label{fig:air_measure_cal}
\end{figure}

The $M_{cal,0,0}$ or top-left element of this matrix corresponds to the power measured as a function of the six free parameters. In practice, by alignment of the polarimeter we can get close to understanding the rotation angles of the system. We should also have some idea of the diattenuation of the polarizers and retardance of the retarders based on the manufacturer's specifications. We use the Newport Agilis series rotation mounts, whose angular steps are unknown. Therefore we also calibrate the angular steps of the mounts in the air measurement. We curve fit to the observed power at the detector plane using the \verb|scipy.optimize.minimize| function with the BFGS optimizer method. The results using 24 measurements are shown in Figure \ref{fig:air_measure_cal}.

\begin{table}
    \centering
    \begin{tabular}{c c c}
        Parameter & Guess Value & Calibrated Value \\
    \hline
         $\theta_{p,2}$ & $90.00^{\circ}$ & $90.00^{\circ}$ \\
         $\theta_{r,2}$ & $15.00^{\circ}$ & $0.08^{\circ}$ \\
         $\delta_{2}$ & $90.00^{\circ}$ & $94.21^{\circ}$ \\
         $\Theta_{2}$ & $10.02^{\circ}$ & $10.36^{\circ}$ \\
         $\theta_{p,1}$ & $0.00^{\circ}$ & $0.00^{\circ}$  \\
         $\theta_{r,1}$ & $-10.00^{\circ}$ & $0.08^{\circ}$ \\
         $\delta_{1}$ & $90.00^{\circ}$ & $92.50^{\circ}$ \\ 
         $\Theta_{1}$ & $9.46^{\circ}$ & $9.34^{\circ}$ \\
    \hline 
    \hline
    \end{tabular}
    \caption{The guess and calibrated free parameters of the DRRP for a sample calibration via measurement of air in Figure \ref{fig:air_measurement_m}. The results of performing this calibration are shown in Figure \ref{fig:air_measure_cal}. The $\Theta$ symbol corresponds to the angular step taken by the rotation stage.}
    \label{tab:my_label}
\end{table}

\begin{figure}
	\centering
	\includegraphics[width=\textwidth]{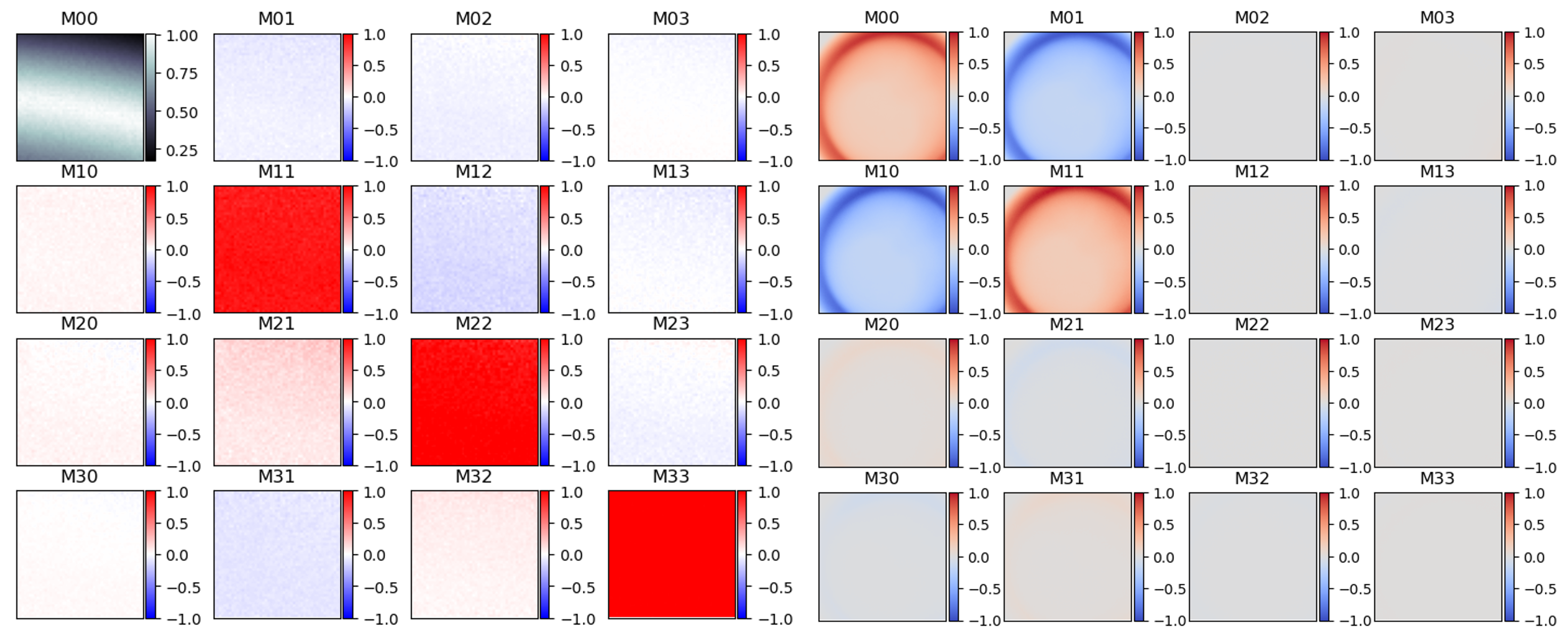}
	\caption{(Left) Spatially-resolved normalized Mueller matrix of air in the presence of nonuniform illumination. The $M_{0,0}$ element is normalized with respect to the maximum power observed, with a non-uniform illumination pattern. The remaining elements of the Mueller matrix images are normalized to the M00 element and appear to be the expected identity matrix with a minimal signal on the off-diagonals. The maximum RMS error to the identity matrix for this measurement is $\approx5\%$ in the $M_{2,1}$ element. (Right) Spatially-resolved non-normalized Mueller matrix image of a vertically oriented linear polarizer in the presence of nonuniform illumination. The $M_{2,1}$ and $M_{3,1}$ images reveal a small amount of retardance, indicating that this the optic under test does not behave like a perfect polarizer.}
	\label{fig:air_measurement_m}
\end{figure}

Fitting the free parameters of our polarimeter to the measured data enables a more accurate understanding of the instrument state, which in turn enables a more accurate polarimetric data reduction matrix. Repeating the same number of measurements with an updated understanding of the instrument state enables a Mueller matrix measurement, which we show in Figure \ref{fig:air_measurement_m} for the case of a simple air (left) and vertical linear polarizer (right) measurement.

\section{Results}
\label{sec:results}

\begin{figure}
    \centering
    \includegraphics[width=0.7\textwidth]{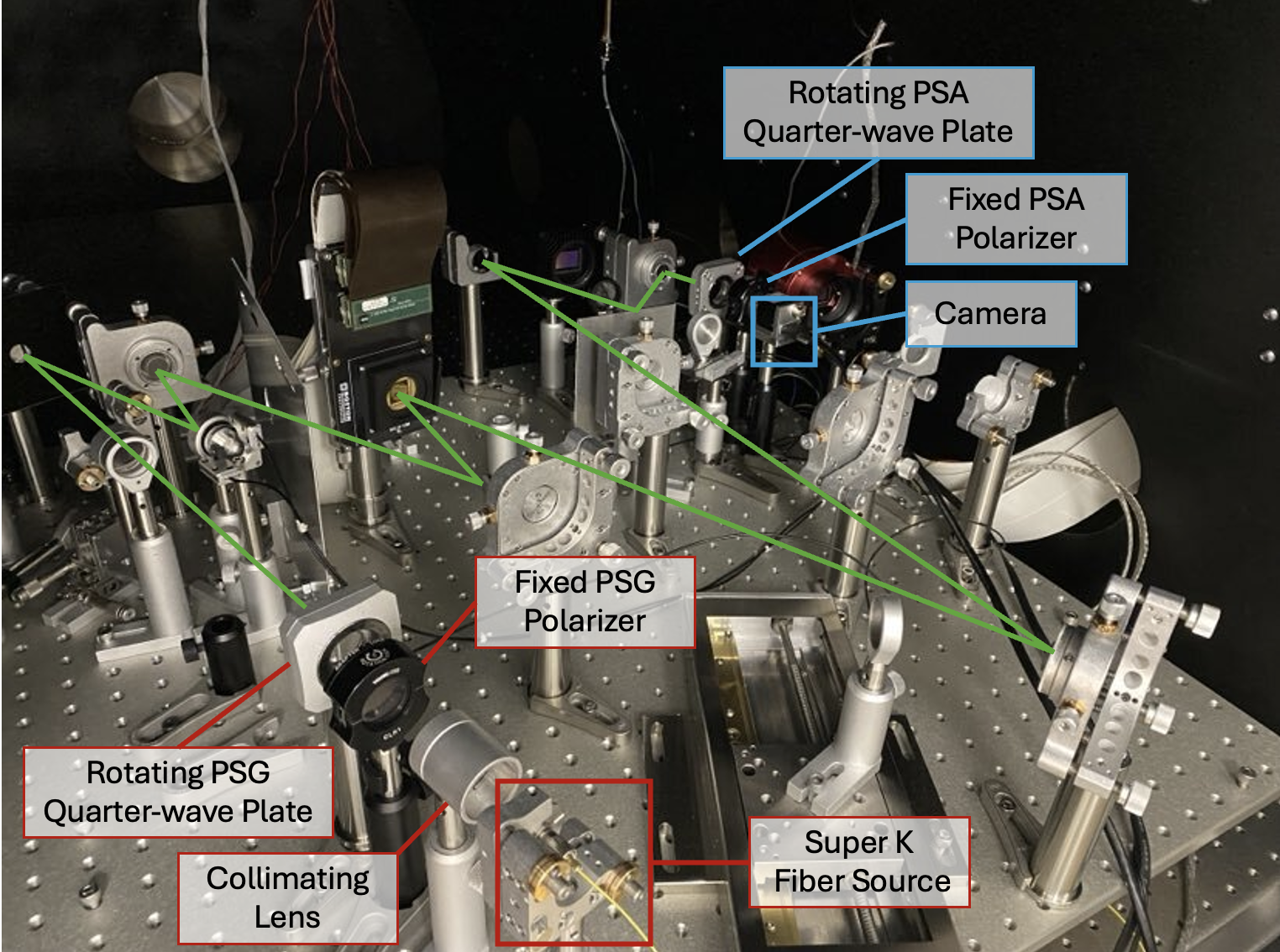}
    \caption{The DRRP Mueller polarimeter assembled around the \scoob testbed. The components of the polarization state generator are denoted in red. The path of the light through \scoob is shown in green. The components of the polarization state analyzer are denoted in blue. Note that along the green path, there are an odd number of (9) reflections.}
    \label{fig:scoob_in_lab}
\end{figure}

\begin{figure}
    \centering
    \includegraphics[width=\textwidth]{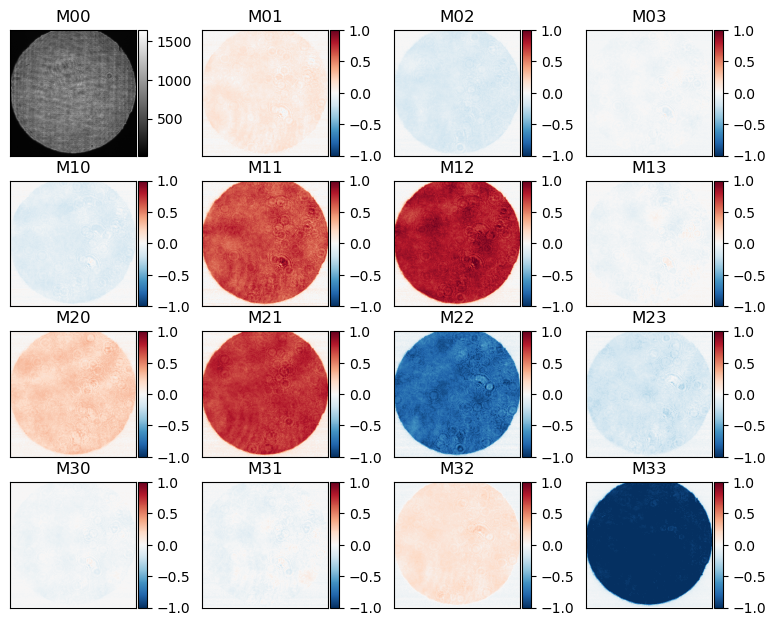}
    \caption{The normalized Mueller Pupil of \scoob at $\lambda=525nm$. The elements of the matrix are normalized element-wise to the M00 element, which is plotted non-normalized in the top-leftmost element to show the pupil illumination. The on-diagonal elements ($M00 \rightarrow M33$) of the Mueller pupil are nearly what one would expect from a system with an odd number of reflections. The most prominent other signals are the $M12$ and $M21$, which represent retardation of Stokes $Q \rightarrow U$ and $U \rightarrow Q$, respectively.}
    \label{fig:mm_525}
\end{figure}

To measure the \scoob Mueller Pupil, we disassemble the polarimeter built in the UASAL laboratory and re-construct it in the TVAC chamber. We then perform an air calibration in the TVAC chamber to calibrate the state of the polarimeter before placing the PSA in the Lyot plane of the coronagraph without the coronagraph mask in, as shown in Figure \ref{fig:scoob_in_lab}. Here a SuperK tunable laser is used for the source instead of the halogen lamp used for the results in Figure \ref{fig:air_measurement_m}. For both the calibration and the measurement of the Mueller Pupil, we take 40 measurements in angular increments of approximately $10^{\circ}$ for the PSG waveplate and $50^{\circ}$ for the PSA waveplate. We take two measurements, one at 630nm and another at 525nm, to understand how the polarization properties change as a function of wavelength. The results for $\lambda=525nm$ are shown in Figure \ref{fig:mm_525}.

While it is useful to consider the full Mueller pupil of \scoob, in order to translate this into a diffraction model, we must separate depolarization from the Mueller pupil. Lu and Chipman famously derived an interpretation of Mueller matrices based on the polar decomposition. In this expression, the Mueller matrix can be considered a multiplication of a depolarizer, retarder, and diattenuator  \cite{Lu:96}, as shown in Equation \ref{eq:polar_decomposition},

\begin{equation}
	\mathbf{M} = \mathbf{M}_{\Delta} \mathbf{M}_{r} \mathbf{M}_{d}.
	\label{eq:polar_decomposition}	
\end{equation}

\begin{figure}
	\centering
	\includegraphics[width=\textwidth]{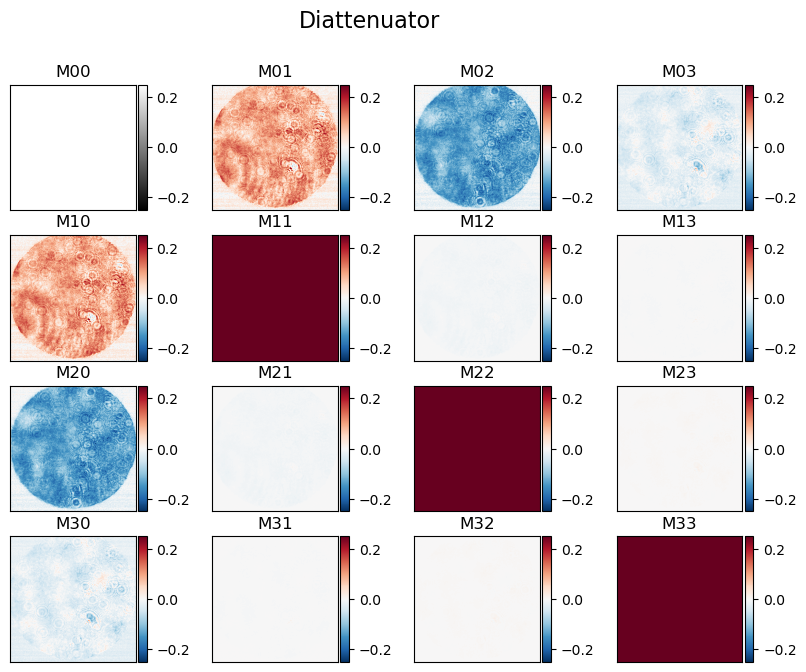}
	\caption{The normalized Diattenuator Mueller Matrix of \scoob from the polar decomposition performed in Figure \ref{fig:mm_525}. The color bar limits are set to highlight the detail in the first row and column on the Mueller matrix. The on-diagonals of this matrix are an array of ones and lie off the color bar scale.}
	\label{fig:mm_525_diattenuator}
\end{figure}

\begin{figure}
	\centering
	\includegraphics[width=\textwidth]{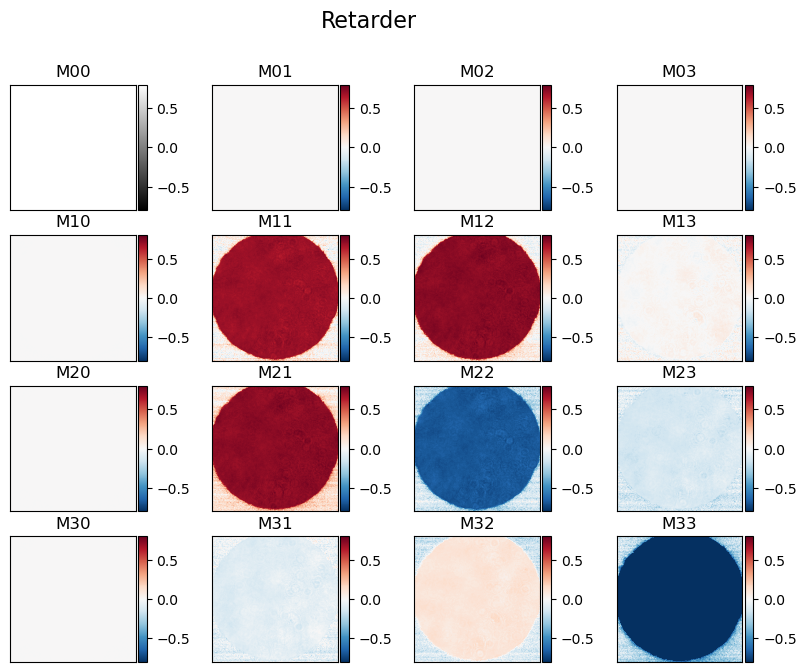}
	\caption{The normalized Retarder Mueller Matrix of \scoob from the polar decomposition performed in Figure \ref{fig:mm_525}. The color bar limits are set to highlight the detail in the bottom-most $3 \times 3$ elements of the Mueller matrix. The M33 element is an array of negative ones and lies off the color bar scale.}
	\label{fig:mm_525_retarder}
\end{figure}

\begin{figure}
	\centering
	\includegraphics[width=\textwidth]{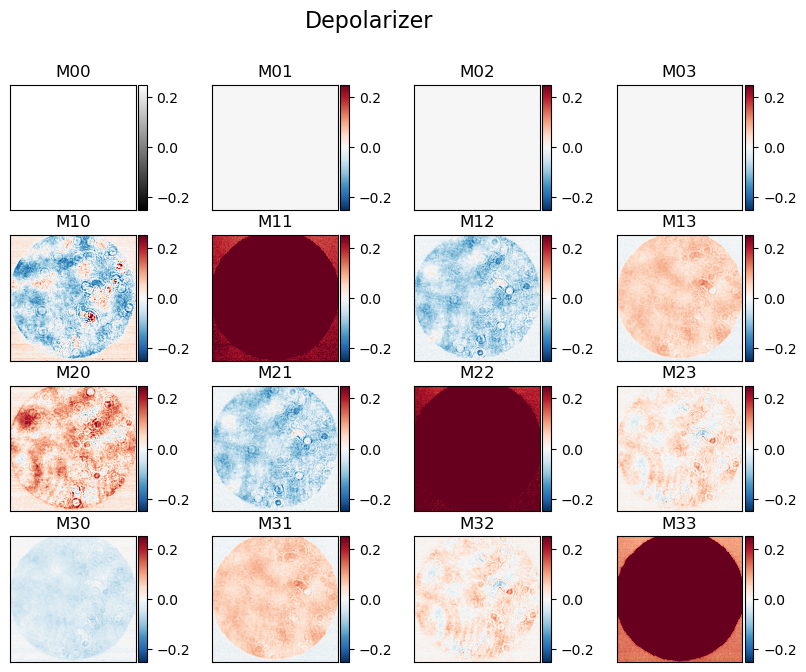}
	\caption{The normalized Depolarizer Mueller Matrix of \scoob from the polar decomposition performed in Figure \ref{fig:mm_525}. The color bar limits are set to highlight the detail in the Mueller matrix. The on-diagonals of this matrix are an array of ones, and lie off the colorbar scale.}
	\label{fig:mm_525_depolarizer}
\end{figure}

Here $\mathbf{M}_{\Delta}$ is the depolarizer, $\mathbf{M}_{r}$ is the retarder, and $\mathbf{M}_{d}$ is the diattenuator. The decomposition method is shown in Lu and Chipman\cite{Lu:96}, but we built the algorithm into the \verb|Katsu| Python package for broader use.  Applying the polar decomposition to the Mueller pupil in Figure \ref{fig:mm_525} yields the data in Figures \ref{fig:mm_525_diattenuator} - \ref{fig:mm_525_depolarizer}. From these data, we can plot the total diattenuation, retardance, and depolarization of the system. The diattenuation ($d$) is given by Equation \ref{eq:diattenuation},

\begin{equation}
	d = \frac{\sqrt{M_{01}^{2} + M_{02}^{2} + M_{03}^{2}}}{M_{00}}.
	\label{eq:diattenuation}
\end{equation} 

The retardance ($\delta$) is given by Equation \ref{eq:retardance},

\begin{equation}
	\delta = cos^{-1}(\frac{Tr(\mathbf{M})}{2} - 1).
	\label{eq:retardance}
\end{equation}

And the depolarization index ($DI$) is given by Equation \ref{eq:depolarization},

\begin{equation}
	DI(\mathbf{M}) = \frac{\sqrt{(\sum_{i,j} M_{i,j}^{2}) - M_{00}^{2}}}{\sqrt{3}M_{00}}.
	\label{eq:depolarization}
\end{equation}

\begin{figure}
	\centering
	\includegraphics[width=\textwidth]{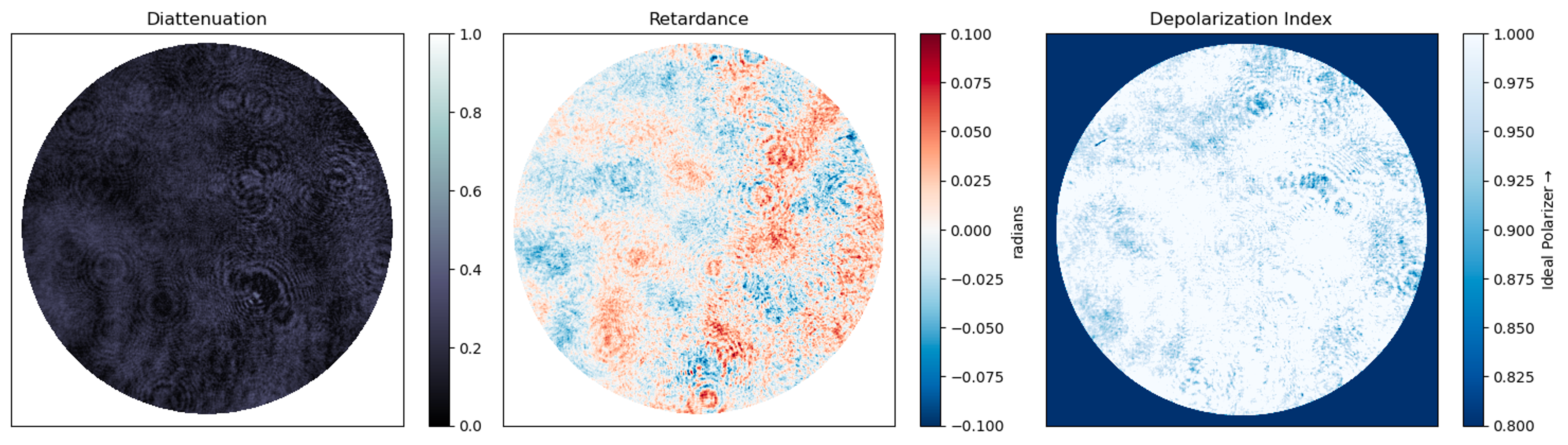}
	\caption{The diattenuation (left), retardance (middle), and depolarization index (right), of the Mueller pupil data for \scoob. The mean value of the diattenuation is $16.5\%$, and the standard deviation of the retardance is $0.03$ radians. The depolarization is very near 1, indicating that there is not a substantial amount of depolarization in the system. Many of the high-frequency structure present in these data is the result of diffraction from dust on the rotating optics and are not indicative of coronagraph performance. However, the mid-frequency structure (4-5 cycles/pupil) indicates polarization aberration from coating nonuniformity or anisotropy.}
	\label{fig:polab_pupils}
\end{figure}

The polar-decomposed Diattenuator, Retarder, and Depolarizer of the measured Mueller pupil in Figure \ref{fig:mm_525} are given in Figures \ref{fig:mm_525_diattenuator}-\ref{fig:mm_525_depolarizer}. Computing the polarization aberration quantities in Equations \ref{eq:diattenuation}-\ref{eq:depolarization} for the Mueller pupil enables us to view the polarization aberrations of the system across the exit pupil of the coronagraph, which we show in Figure \ref{fig:polab_pupils}.  There is a diattenuation across the pupil of approximately $16.5\%$ with a standard deviation of $7.3\%$. Furthermore, we observe nonuniform behavior in the retardance plot that has a standard deviation of approximately $0.03$ radians, or $\approx \lambda / 200$. The depolarization index across the pupil is very near 1, indicating a lack of depolarization in the system. The standard deviation of these data is $\approx 5\%$, but much of that signal is dominated by artifacts from the dust diffraction, so the depolarization of the coronagraph should be less than that. 

The mid-frequency structure that is apparent in Figures \ref{fig:mm_525} - \ref{fig:polab_pupils} is perhaps the most interesting result of this study.   They are of a higher spatial order than is predicted by typical polarization ray traces of simple systems (tilt/focus-like shapes), and consequently, their origin is unknown. Note that these patterns are not present in the laboratory measurements of air and a linear polarizer in Figure \ref{fig:air_measurement_m}. Most likely, we are observing the influence of imperfect coatings on the polarized exit pupil of the coronagraph. Understanding how these effects translate to the minimum achievable contrast of the coronagraph will be critical to assessing how sensitive coronagraphs are to coating nonuniformity. 

For a preliminary analysis into the impact of the as-measured polarization aberrations on \scoob's contrast floor, we can convert the non-depolarizing Mueller pupil into a Jones pupil and propagate it through a charge-6 vector vortex coronagraph using HCIPy\cite{por2018hcipy}. The computation of the Jones pupil is illustrated by Equation \ref{eq:mul_to_jones},

\begin{equation}
    \mathbf{J} = 
    \begin{pmatrix}
        A_{xx} & A_{xy}e^{i(\phi_{xy}-\phi_{xx})} \\ 
        A_{yx}e^{i(\phi_{yx}-\phi_{xx})} & A_{yy}e^{i(\phi_{yy}-\phi_{xx})} \\ 
    \end{pmatrix}.
    \label{eq:mul_to_jones}
\end{equation}

\begin{figure}
    \centering
    \includegraphics[width=\textwidth]{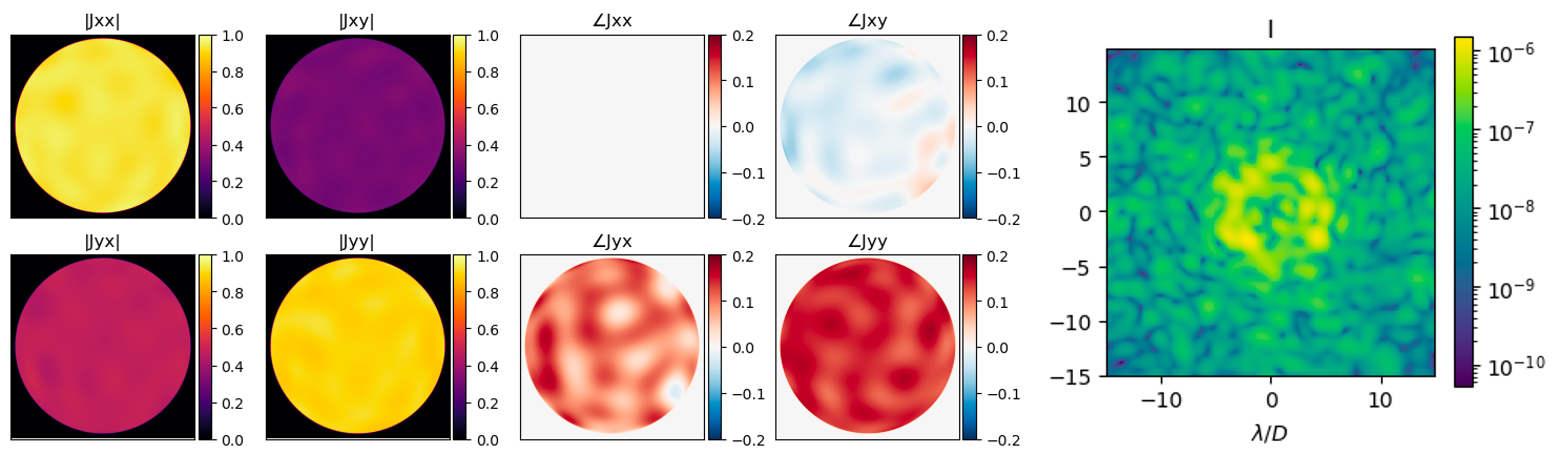}
    \caption{(Left) The Jones pupil computed from the Mueller pupil in Figure \ref{fig:mm_525} using Equation \ref{eq:mul_to_jones}. We low-pass filter the Jones pupil by representing it as the first 150 Noll-indexed Zernike polynomials so that the artifacts from the dust diffraction are removed from the data. The Jones pupil is then propagated through a charge-6 vector vortex coronagraph, which is shown in units of normalized intensity on the right. The mid-spatial frequency errors manifest as residuals of $\approx 10^{-6}$ normalized intensity from $4-5\lambda/D$. Note that these data are uncompensated, and are not representative of the focal plane after wavefront control. As shown in Van Gorkom et al\cite{VanGorkom2024}, \scoob is presently capable of getting to $10^{-8}-10^{-9}$ contrast from $3-10\lambda/D$. Therefore the influence of these polarized ghosts subject to wavefront control algorithms needs to be studied in future work.}
    \label{fig:jones_525}
\end{figure}

Where the amplitudes are given by Equations \ref{eq:Axx}-\ref{eq:Ayy}, and the relative phases are given by Equations \ref{eq:phixy}-\ref{eq:phiyy}. Note that while the amplitudes are completely determined, the phases must be determined relative to some reference because global phase is lost when measuring intensity. The Jones matrix and corresponding focal plane after propagation through a charge-6 vector vortex coronagraph with an $80\%$ diameter Lyot stop is shown in Figure \ref{fig:jones_525}. The as-measured polarization aberrations appear to result in a degradation of the normalized intensity from $4-5\lambda/D$, but as shown in Van Gorkom et al (these proceedings)\cite{VanGorkom2024}, after the application of focal plane wavefront sensing and control, \scoob is capable of reaching contrasts of $\approx10^{-8}-10^{-9}$. Therefore, we know that these polarized wavefront errors are not fundamentally limiting at these locations on the focal plane. In future work we will assess the influence of these as-measured polarization aberrations on a simulated \scoob dark hole in the presence of wavefront sensing and control.

\begin{equation}
    A_{xx} = \frac{\sqrt{M_{00} + M_{01} + M_{10} + M_{11}}}{2}
    \label{eq:Axx}
\end{equation}
\begin{equation}
    A_{xy} = \frac{\sqrt{M_{00} - M_{01} + M_{10} - M_{11}}}{2}
\end{equation}
\begin{equation}
    A_{yx} = \frac{\sqrt{M_{00} + M_{01} - M_{10} - M_{11}}}{2}
\end{equation}
\begin{equation}
    A_{yy} = \frac{\sqrt{M_{00} - M_{01} - M_{10} + M_{11}}}{2}
    \label{eq:Ayy}
\end{equation}
\begin{equation}
    \phi_{xx} - \phi_{xy} = tan^{-1}(\frac{M_{03} + M_{13}}{M02 + M12})
    \label{eq:phixy}
\end{equation}
\begin{equation}
    \phi_{yx} - \phi_{xx} = tan^{-1}(\frac{M_{30} + M_{31}}{M20 + M21})
\end{equation}
\begin{equation}
    \phi_{yy} - \phi_{xx} = tan^{-1}(\frac{M_{32} + M_{23}}{M22 + M33})
    \label{eq:phiyy}
\end{equation}

\section{Discussion}
\label{sec:discussion}
This study introduces the dual-rotating-retarder Mueller polarimeter as a characterization methodology for assessing the polarized performance of coronagraphic optical systems. We built our methods and data reduction into an open-source Python package called \verb|katsu|, which is freely available on GitHub. Using \verb|katsu|, we illustrate the principle of dual-rotating-retarder Mueller polarimetry in the lab for simple optical element characterization and then apply it to the polarimetry of the \scoob testbed. We find a nominal diattenuation of $16.5\%$, and a retardance standard deviation of $\pm \lambda / 200$, with little depolarization. We also observe a clear mid-frequency (4-5 cycles/pupil) structure in each of the elements of the Mueller pupil and corresponding polarization aberration pupils in Figure \ref{fig:polab_pupils}. Understanding the influence of this aberration will be the subject of future work to calibrate the static contrast floor of \scoob. 

Understanding the origin of the mid-frequency aberration revealed in this study is a key next step for this work. However, characterizing film uniformity across an aperture is a non-trivial task. In principle, this could be achieved through ellipsometry. By depositing the same mirror coating used for the \scoob mirrors on a flat substrate, the ellipsometric characteristics can be studied at multiple points across the substrate. These data can be fit to a layer model that includes a refractive index tensor to more precisely understand the degree of birefringence in the system. 

\acknowledgments 
 
This work was supported by a NASA Space Technology Graduate Research Opportunity. The research presented in these proceedings was predominantly conducted with open-source software, including: Poke\cite{Ashcraft_2023}, Katsu\cite{Ashcraft_Katsu}, HCIPy\cite{por2018hcipy}, numpy\cite{harris2020array}, scipy\cite{2020SciPy-NMeth}, matplotlib\cite{Hunter:2007}, and IPython\cite{PER-GRA:2007}.

\section{Appendix:}

\begin{figure}
	\centering
	\includegraphics[width=\textwidth]{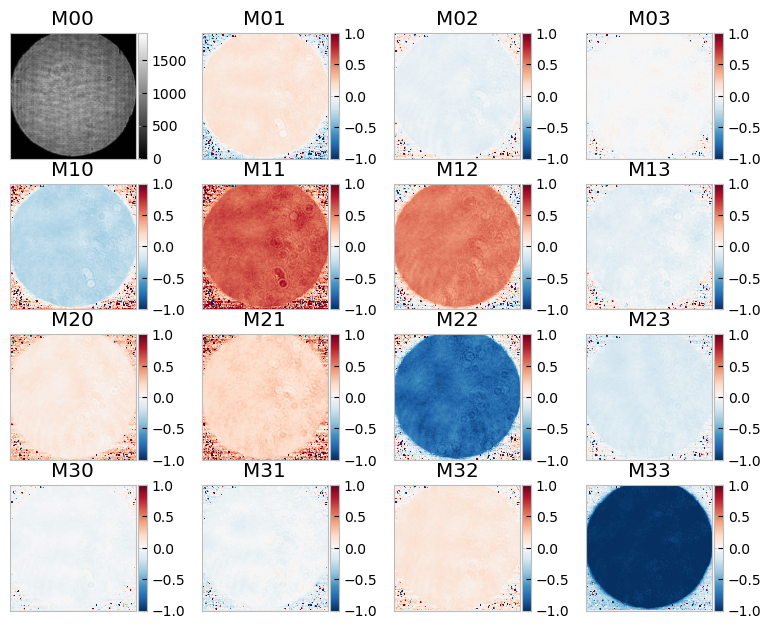}
	\caption{The normalized Mueller Pupil of \scoob at $\lambda=630nm$. The elements of the matrix are normalized element-wise to the M00 element, which is plotted non-normalized in the top-leftmost element to show the pupil illumination. The on-diagonal elements ($M00 \rightarrow M33$) of the Mueller pupil are nearly what one would expect from a system with an odd number of reflections. This Mueller Matrix image reveals a greater linear diattenuation in the first row of the Mueller matrix and less retardance that couples Stokes $Q \rightarrow U$ and $U \rightarrow Q$, respectively.}
	\label{fig:enter-label}
\end{figure}

\begin{figure}
    \centering
    \includegraphics[width=\textwidth]{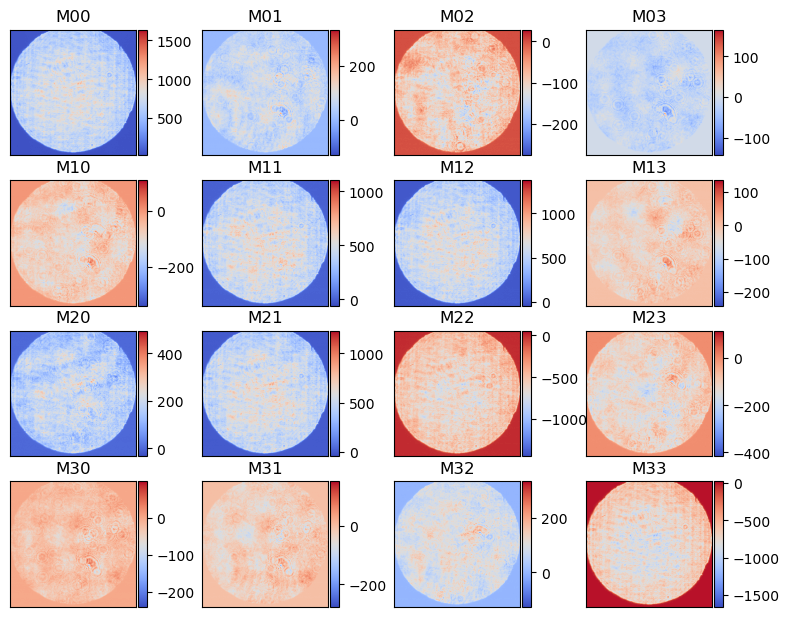}
    \caption{Un-normalized Mueller matrix at $\lambda=525nm$. }
    \label{fig:enter-label}
\end{figure}

\begin{table}[H]
    \centering
    \begin{tabular}{c c}
    \hline
        Vendor  & Part \\
    \hline
        Thorlabs & \href{https://www.thorlabs.com/thorproduct.cfm?partnumber=SLS201L}{Stabilized Tungsten-Halogen Light Source SLS201L} \\ 
        NKT Photonics & \href{https://www.nktphotonics.com/products/supercontinuum-white-light-lasers/superk-compact/}{SuperK COMPACT white light laser} \\
        Newport & \href{https://www.newport.com/p/AG-PR100V6}{Agilis PR100-V6} \\
        Newport & \href{https://www.newport.com/p/AG-UC8}{Agilis AG-UC8 Controller} \\
        Meadowlark & \href{https://photonics.laser2000.co.uk/wp-content/uploads/2021/08/PPM.pdf}{High-contrast Linear Polarizer} \\
        BVO & \href{https://boldervision.com/waveplates/aqwp2/}{Quarter-wave Plate} \\
    \hline
    \hline
    \end{tabular}
    \caption{Table of critical components used in the polarimeter assembled in UASAL. Additional mounting hardware available from Thorlabs and Newport were used to mount the optics, but the precise models are not critical to this study.}
    \label{tab:my_label}
\end{table}

\bibliography{report} 
\bibliographystyle{spiebib} 

\end{document}